\documentclass[prb,aps,draft]{revtex4}
\input epsf

\begin{document}
\title{
Ballistic trajectory: parabola, ellipse, or what?}
\author{Lior M. Burko}
\author{Richard H. Price}
\altaffiliation[New address: ]{Department of Phyiscs and Astronomy
and Center for Gravitational Wave Astronomy, 
The University of Texas as Brownsville, Brownsville,
Texas, 78520.
}\affiliation{Department of Physics, University of Utah, Salt Lake City, Utah
84112.}

\begin{abstract}
Mechanics texts tell us that a particle in a bound orbit under
gravitational central force moves on an ellipse, while introductory
physics texts approximate the earth as flat, and tell us that the
particle moves in a parabola. The uniform-gravity, flat-earth parabola is
clearly meant to be an approximation to a small segment of the true
central-force/ellipse orbit. To look more deeply into this connection
we convert earth-centered polar coordinates to ``flat-earth
coordinates'' by treating radial lines as vertical, and by treating
lines of constant radial distance as horizontal. With the exact
trajectory and dynamics in this system, we consider such questions as
whether gravity is purely vertical in this picture, and whether the
central force nature of gravity is important only when the height or
range of a ballistic trajectory is comparable to the earth
radius. Somewhat surprisingly, the answers to both questions is
``no,'' and therein lie some interesting lessons.
\end{abstract}

\maketitle

\section{Introduction}\label{sec:intro} 
\begin{figure}[h]
  \begin{center}
{\epsfxsize=350pt \epsfbox{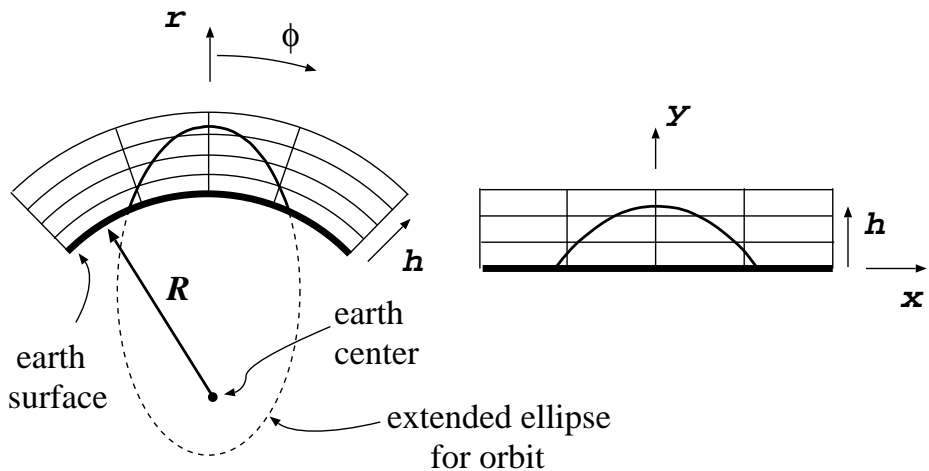}}
  \end{center}
\caption{\label{fig:coords} Ellipse for a ballistic trajectory
pictured in polar coordinates and in flat earth coordinates.}
\end{figure}

The trajectory of a particle moving without drag under the influence
of the gravitational field of a perfectly spherical earth is a conic
section. Since a cannonball, for example, is on a bound orbit, its
flight from muzzle to target is part of an ellipse, as shown on the
left in Fig.~\ref{fig:coords}. The center of the earth is at the
(distant) focus of the highly eccentric ellipse, and the part of the
ellipse relevant to cannonball flight is a small segment near the
apogee. Introductory texts, however, treat ballistic motion using
Galileo's approach. Specifically, the horizontal and vertical
components of the motion of the projectile are separated, and the
trajectory is given by a parabola. The parabola as a limit of an
ellipse was noted by Newton in the {\em Principia}\cite{principia}:
\begin{quote}
If the ellipsis, by having its centre removed to an infinite distance,
degenerates into a parabola, the body will move in this parabola; and the
force, now tending to a centre infinitely remote, will become
equable. Which is {\it Galileo's} theorem. 
\end{quote}

Many of the introductory textbooks we have reviewed do not discuss the
nature of the approximation, i.e., they do not discuss the conditions
under which the flat-earth parabola
is an accurate approximation
for what is really a central force ellipse. Some textbooks,
however, do make explicit statements about the approximation,
statements that seem plausible, but turn out to be incorrect.
Specifically, we found three classes of conditions which are used in those
textbooks that do discuss the validity of the approximation. In the first
class it is assumed that the maximal height of the trajectory above the
surface of the earth is small compared with the radius of the
earth (or, equivalently, that the {\em magnitude} of the
gravitational acceleration does not change by much along
the trajectory)\cite{height}; in the second class
it is assumed that the range of the trajectory is small compared with the
radius of the earth (or, equivalently, that the {\em direction} of the 
gravitational acceleration does not change by much along the
trajectory)\cite{range};  in the third class it is assumed that
both the height and the range are small compared with the radius of the
earth (both the magnitude and the direction of the gravitational
acceleration do not change by much)\cite{height_and_range}. Other
textbooks refer vaguely to
neglecting
the curvature of the earth in justifying the approximation, but do not
quantify the condition \cite{vague}.

In this paper we show that the correct condition under which the
flat-earth picture is a good approximation for the trajectory, is that
the maximal curvature of the trajectory is much greater than the
curvature of the surface of the earth. This condition coincides with
the three classes of naive conditions for many, but not all,
trajectories. There are trajectories for which all the three
naive classes of conditions are satisfied, but for which the
approximation fails to be valid.

To ``flatten the earth'' without any approximation, we introduce
geocentric polar coordinates $r,\phi$ in the plane of the orbit (as
shown on the left of Fig.~\ref{fig:coords}) and we plot them (on the
right in Fig.~\ref{fig:coords}) as if they were Cartesian coordinates
$x,y$. We let $R$ denote the radius of our perfect earth, and we take
$h$ to be the height, above the earth, of the peak of the
trajectory. The apogee of the particle orbit is then at a distance
$r=R+h$ from the center of the earth.  For convenience, we choose our
zero of $\phi$ so that it coincides with the apogee of the orbit.
In order to view the space above the earth in this way, as
if the earth were flat, we need to distort the coordinates in somewhat
the same manner as a map maker.  We choose to do this with the explicit
relationship \cite{notunique}
\begin{equation}\label{coordxform} 
x=(R+h)\phi\quad\quad\quad\quad y=r-R\ .
\end{equation}
and to plot $x,y$ as if they were Cartesian coordinates. 

In this paper we shall focus on two questions about exact ballistic
trajectories. The first is the accuracy of the parabolic approximation.  
The naive expectation of most students (and many teachers) is
that the approximation is valid if the trajectory is ``small'' compared to
the radius of the earth, i.e.\,, if neither $h$ nor the range of the
trajectory
is comparable to the earth radius $R$. We shall
see that this naive expectation is too naive.

Our second question involves the vertical nature of gravitation.  In
the ``true'' picture of gravitational acceleration, on the left side
of Fig.~\ref{fig:coords}, the gravitational force is radially directed
towards the center of the earth. These radial lines are converted to
vertical lines by the redrawing of the coordinates on the right hand
side of Fig.~\ref{fig:coords}, and by the transformation in
Eq.~(\ref{coordxform}). It would seem, therefore, that gravitational
acceleration {\em must} be acting vertically in the picture on the
right hand side of Eq.~(\ref{coordxform}). The introductory textbooks
add to this the assumption that $g$ is constant, and prove that a
particle being acted upon only by gravitational forces moves in a
parabola. This argument leads unavoidably to the conclusion that the
particle trajectory deviates from a parabola only due to the slight
variation of $g$ with altitude. We shall see that again the 
naive expectation is not correct.

\section{Elliptical orbits in flat-earth coordinates}\label{sec:ellip} 
To describe the orbit of a unit mass particle we start by letting
\begin{displaymath}
L=r^2\frac{d\phi}{dt}
\end{displaymath}
represent the particle angular momentum per unit particle mass. With standard
techniques\cite{anymechbook}  we can
write the equation of the particle's orbit, in polar coordinates, as
\begin{equation}\label{ellipseq} 
r=\frac{1}{\frac{GM}{L^2} +\left(\frac{1}{R+h}
-\frac{GM}{L^2}\right)\cos{\phi}}\ ,
\end{equation}
which is the equation of an ellipse (though not in the most familiar 
form). Here $M$ is the mass of the earth and $G$ is the universal gravitational
constant.

We can convert Eq.~(\ref{ellipseq}) to the flat-earth picture
simply by introducing the $x,y$ coordinates of Eq.~(\ref{coordxform}), 
and can write the trajectory in the right hand side of Fig.~\ref{fig:coords}
as
\begin{equation}\label{trajeq} 
y+R=\frac{1}{\frac{GM}{L^2} 
+\left(\frac{1}{R+h}-\frac{GM}{L^2}\right)\cos{(x/[R+h])}}\ .
\end{equation}
To simplify the complex appearance of this equation we introduce two
dimensionless parameters. The first is $\epsilon\equiv h/R$, a parameter
that tells us something about the relative size of the trajectory and
the earth. We want our second parameter to be expressed in terms of
quantities appropriate to the usual description of a ballistic
trajectory. To that end we introduce the parameter $V_{horiz}$, the
particle velocity at the top of the trajectory, and we combine it with
the acceleration of gravity at the surface of the earth $g$
to form the dimensionles  parameter
\begin{equation}\label{alphadef} 
\alpha\equiv\frac{V^2_{horiz}}{gh}\ .
\end{equation}
We next note that $V_{horiz}=L/(R+h)$ and $g=GM/R^2$, so that $\alpha$
is equivalent to $(L^2/GMh)/(1+\epsilon)^2$. With these equivalences we can
rewrite 
the trajectory  in Eq.~(\ref{trajeq}) as
\begin{equation}\label{exact} 
\frac{y}{h}=\frac{\alpha(1+\epsilon)^2}{1+
\left[\alpha\epsilon(1+\epsilon)-1\right]
\cos{
\left(\frac{\epsilon(x/h)}{[1+\epsilon]}\right)
}
}-\frac{1}{\epsilon}\ .
\end{equation}

\begin{figure}[h]
  \begin{center}
{\epsfxsize=250pt \epsfbox{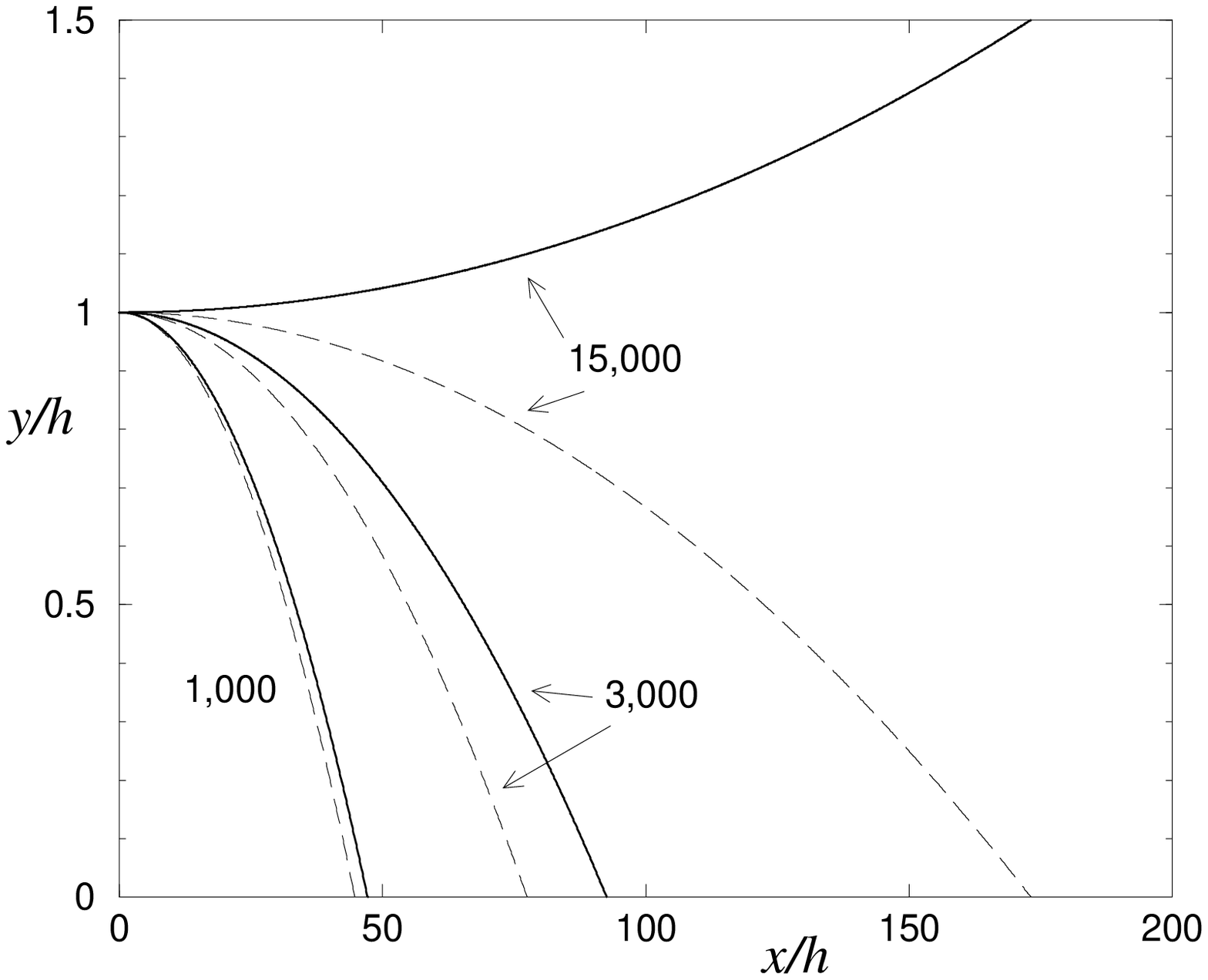}}
  \end{center}
\caption{\label{fig:largealph}}
Trajectories for $h/R=10^{-4}$.
Solid curves are exact trajectories in flat-earth coordinates
and dashed curves are parabolic approximations. Curves are labeled 
by the value of $\alpha$ to which they correspond.
\end{figure}

In Eq.~(\ref{exact}), the only length scale that explicitly appears is
$h$, a scale appropriate to the description of ballistic trajectories. But the parameter
$\epsilon\equiv h/R$ still carries information about the size of the
earth.  The familiar parabolic trajectory
follows when this parameter is taken to be very small. To see this, we 
can expand the right hand side in powers of $\epsilon$ and keep only
the term to zero order in $\epsilon$ 
(keeping $\alpha$ fixed). The result is
\begin{equation}\label{standard} 
\frac{y}{h}=1-\frac{1}{2\alpha}\frac{x^2}{h^2}
=1-\frac{g}{2V^2_{horiz}}\frac{x^2}{h}\ ,
\end{equation}
the standard parabolic ballistic trajectory,

We now look more carefully at the exact orbit in the flat-earth
coordinates of Eq.~(\ref{exact}). In Fig.~\ref{fig:largealph} we
present trajectories for $\epsilon=10^{-4}$, hence for
$h\approx0.63$\,km.  The derivation above, of the parabolic
approximation, assumes that $\alpha$ is of order unity.  To see
interesting deviations from a parabola, we consider large values of
$\alpha$ in the figure.  For $\alpha=15,000$, the deviations from the
parabola are striking indeed; the trajectory bends upward. This, of
course, is simply the appearance in our coordinates of an elliptical
orbit for which the radius of curvature at the apogee is greater 
than the radius of curvature of the earth's surface. 

What may be most surprising in Fig.~\ref{fig:largealph}, is that large
deviations from the parabola occur when the range of a trajectory is
only around $90h$, or about 60\,km, a tiny distance compared to the
size of the earth.  To see why this is so, we can put $y=0$ in
Eq.~(\ref{exact}) and we can solve for $x$, half the range of the
particle:
\begin{equation}
\frac{x}{h}=\frac{1+\epsilon}{\epsilon}\cos^{-1
}{\left(\frac{1-\epsilon\alpha(1+\epsilon)^2}
{1-\epsilon\alpha(1+\epsilon)}\right)}
=\sqrt{2\alpha}\left[
1+\frac{\epsilon}{2}(\alpha-1)
\right]+{\cal O}\left(\epsilon^2\right)\ .
\end{equation}
The fractional error in  the range of the trajectory, for large $\alpha$ is
therefore 
\begin{equation}\label{range} 
\frac{\epsilon}{2}(\alpha-1)\approx\frac{\epsilon\alpha}{2}
\approx\frac{{\rm range}^2}{16Rh}\ ,
\end{equation}
where we have used the fact that the range is approximately
$2h\sqrt{2\alpha}$. Equation (\ref{range}) shows that the deviation
from the standard range formula is
large when the range is of order $\sqrt{Rh}$ or $R\sqrt{\epsilon}$,
and hence can be important for orbits that are much smaller than the
scale of the earth\cite{loworbit}. This conclusion is in good agreement with 
the $\alpha=3000, \epsilon=10^{-4}$ example in
Fig.~\ref{fig:largealph}. For these parameters the parabolic approximation
misses the target by $15\%$. 

There is a nice way of understanding what we have just found.  The
replacement of the spherical earth surface with a flat surface, in
Fig.~\ref{fig:coords}, is justifiable only if the trajectory is more
sharply curved than is the earth surface.  That is, if the minimal radius
of curvature along the trajectory is smaller than the radius of the
earth. The radius of curvature\cite{curvature} ${\cal R}$ of the
trajectory is\cite{curvature_ellipse}
\begin{equation}\label{curvature_parabola}
{\cal R} := 
\frac{\left[1+(dy/dx)^2\right]^{3/2}
}{\left|d^2y/dx^2\right|}
=\frac{h\alpha}{\left[1+x^2/
\left(h\alpha\right)^2
\right]^{3/2}
}\ ,
\end{equation}
where we have used Eq.~(\ref{standard}). The condition that the
maximum curvature of the trajectory (the curvature at $x=0$) is
much greater
than the curvature of the earth's surface is that $h\alpha/R\ll 1$. This
is violated when $\epsilon\alpha\sim 1$ for which we have, from
Eq.~(\ref{range}), a large deviation from the parabolic range
prediction. We can now
understand this in terms of radius of curvature of the trajectory. For a
small $h$, high velocity orbit, a projectile can move ``just above'' the
surface of the earth on a trajectory with a range much larger than its
height $h$ (but much smaller than $R$).  The curvature of that trajectory
in space is mostly due to the curvature of the earth. In the flat earth
picture the true trajectory would seem to have almost constant height, and
would greatly deviate  from a parabola\cite{latus_rectum}.
It should be noted that the meaning of a flat earth trajectory
disappears if $\epsilon\alpha>1$. It follows from
Eq.~(\ref{curvature_parabola}) that for $\epsilon\alpha>1$ the radius
of curvature at the apogee exceeds that of the earth. This suggests
that the elliptical or hyperbolic
trajectory of the projectile will not intersect the earth
surface. That this is, in fact, true can be verified by looking for
solutions of $y=0$ in Eq.~(\ref{exact}); for $\epsilon\alpha>1$, there
are no solutions with positive $\epsilon$.

There is also an interesting non-geometrical way of viewing the 
condition $\alpha\epsilon\ll1$. From the definitions of $\alpha$, $\epsilon$,
and $g$, we have
\begin{equation}\label{lowveloc} 
\alpha\epsilon=\frac{V^2_{horiz}}{gR}=
\frac{V^2_{horiz}}{V^2_{circ}}=\frac{2V^2_{horiz}}{V^2_{esc}}
\,,
\end{equation}
where $V_{circ}$ is the velocity of a circular orbit with radius $R$
(i.e.\,, just above the earth surface), and $V_{esc}$ is the escape
velocity from the earth surface. The condition
$\alpha\epsilon\ll1$, then, is simply the
condition that the motion is very slow compared to typical ``orbital''
(as opposed to ``trajectory'') motions.  For trajectories with
$V_{horiz}$ comparable to $V_{esc}$ we should not be surprised that
the deviation from a parabolic projectile trajectory is
significant. We {\em may}, however, be surprised that the range of
that trajectory may be short, much less than the earth radius. The key
idea here is that the range is kept short by imposing a small value of
$h$.  For sufficiently small values of $h$ the high velocity
projectiles hit the earth's surface before they get a chance to show
how far they would have gone in their large elliptical orbits. In
order for this to happen, in order for the range to be kept small, the
value of $h$ must be reduced as $V_{horizon}$ is increased.  The
limiting case is $V_{horizon}=V_{esc}/\sqrt{2\;}$, the maximum for which
the trajectory has a range (i.e., for which the trajectory intercepts
the earth's surface). As this limit is approached, the condition on
$h$, for the range to be much less than the earth radius, is
$\epsilon\ll\left[1-2(V_{horizon}/V_{esc})^2\right]$.


%
\section{Is gravity vertical in the flat-earth picture?}\label{sec:vertical}  

In Section \ref{sec:intro} we argued that gravitational acceleration must be
vertical in the flat earth picture. 
It is apparent in Fig.~\ref{fig:largealph} that this cannot be
correct.  For the $\alpha=15,000$ trajectory, the particle is
accelerating upward.  And this particle is at a height of only a
kilometer or so above the earth's surface, where the acceleration of
gravity is certainly close to 9.8\,m/sec$^2$ and is even more
certainly {\em downward}.

\begin{figure}[h]
  \begin{center}
{\epsfxsize=250pt \epsfbox{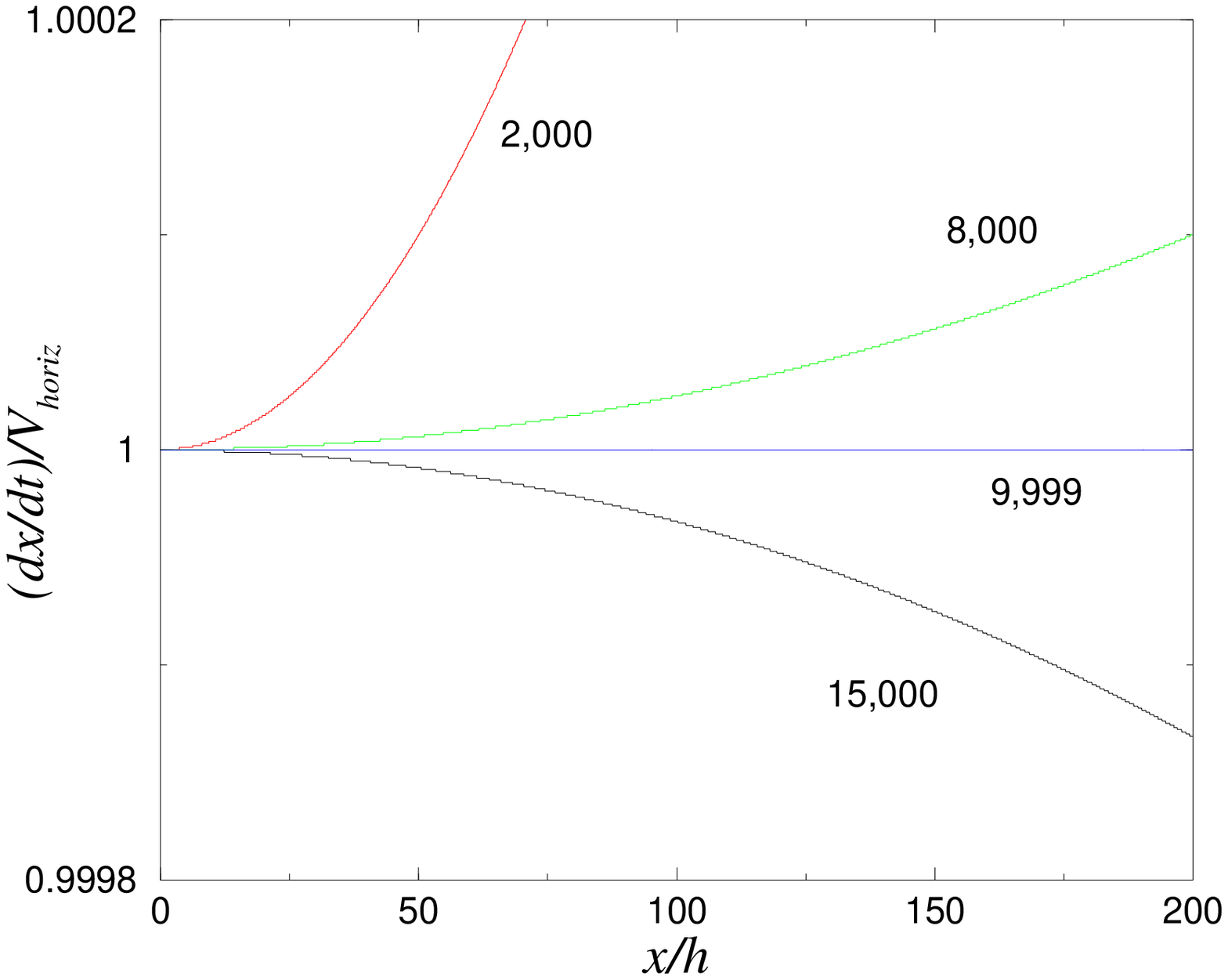}}
  \end{center}
\caption{\label{fig:veloc}}
Trajectories for $h/R=10^{-4}$.
The horizontal velocity $dx/dt$ is shown as a function of horizontal
position $x$. ($V_{horiz}$ is the horizontal velocity at the apogee.) 
Curves are labeled by $\alpha$. The value $\alpha=9,999$
represents a circular orbit.
\end{figure}

Another clear indication that ``vertical acceleration'' is not 
the whole story, is the fact that the horizontal velocity $dx/dt$
changes in time. Figure \ref{fig:veloc} shows the small, but nonzero,
fractional change in the horizontal velocity along the orbit.
To understand the results in this figure we note that 
\begin{equation}\label{dxdt} 
\frac{1}{V_{horiz}}\frac{dx}{dt}=\frac{(R+h)}{V_{horiz}}\frac{d\phi}{dt}=\frac{(R+h)L}{V_{horiz}r^2}=\frac{(R+h)^2}{r^2}\ .
\end{equation}
In this simple calculation it is easy to see why the horizontal
velocity changes: it is proportional to $L/r^2$. The angular momentum
$L$ is constant, but $r$ is not constant except in the case of the
circular orbit for which $\alpha=\alpha_{crit}\equiv R^2/h(R+h)$.  As
the particle moves to smaller radius (in the case $\alpha<
\alpha_{crit}
$) the horizontal velocity must increase; as the particle moves
to larger radius (in the case $\alpha>\alpha_{crit}$) the particle velocity
must decrease.  In this mathematics it is easy to see why the
horizontal velocity changes during the motion.

At a deeper level, the failure of naive intuition is due to the 
tacit expectation that the equations of motion for a unit mass particle
can be put in a form 
\begin{equation}\label{intuition} 
\frac{d^2x}{dt^2}=F_x\quad\quad\quad\frac{d^2y}{dt^2}=F_y\,,
\end{equation}
where $F_x$ and $F_y$ are only functions of position $x,y$. 
Since particles released from rest will fall vertically downward
we conclude that $F_x=0$, and gravity is vertical.

We are misled to this conclusion by the Cartesian appearance 
of the $x,y$ coordinates on the right side of Fig.~\ref{fig:coords}.
The actual equations of motion\cite{eqsofmotion},  turn out to be
\begin{eqnarray}
\frac{d^2x}{dt^2}&=&-\frac{2}{(y+R)}\frac{dx}{dt}\frac{dy}{dt}
\label{trueaccel_x}
\\
\frac{d^2y}{dt^2}&=&-\frac{GM}{(y+R)^2}
+\frac{(y+R)}{(R+h)^2}\left(\frac{dx}{dt}\right)^2\, .
\label{trueaccel}
\end{eqnarray}
In the case of a vertical trajectory ($dx/dt=0$) the equations are in
complete agreement with our expectations. When $dx/dt\neq0$, however,
the velocity dependent terms, absent in Eqs.~(\ref{intuition}), 
are responsible for the failure of intuiton\cite{notG,neglect_vel}. 
These velocity
dependent terms are due to the fact that the $x,y$ coordinate lines are
actually curved lines through physical space. We can draw them as straight
lines, but the way in which we draw them, or the symbols we use to
represent them, do not change the fact that they are curved in physical
space. This curvature of spatial coordinates comes with its usual 
kinematical consequences. The $(dx/dt)^2$ in Eq.~(\ref{trueaccel}),
for example, is just the usual centripetal term disguised by unusual
coordinate names.

\section{Conclusions}\label{sec:conc} 
It is plausible to expect that the uniform-gravity, flat-earth gravity
parabola is an accurate approximation if the height $h$ of the
trajectory, and the range of the trajectory, are both much less than
the earth radius $R$. In terms of our parameterization these
conditions are, respectively $\epsilon\ll1$ and
$\alpha\epsilon^2\ll1$.  If we are to have the kinematics correct, as
well as the shape of the orbit, then clearly the first condition is
necessary; if $h$ is not small compared to $R$ gravity cannot be of
uniform strength.

The second condition, however, is not sufficiently strict. The correct
second condition for the validity of the flat-earth approximation is
that the maximum curvature of the trajectory be much greater than the
curvature of the earth's surface. In our notation this corresponds to
orbits that satisfy $\alpha\epsilon\ll1$. This condition is not
satisfied for a class of trajectories that {\em do} satisfy $\epsilon\ll1$
and $\alpha\epsilon^2\ll1$. That class consists of low height, high
velocity motions. The importance of the high velocity feature of these
orbits can be seen in the fact that our condition $\alpha\epsilon\ll1$
is equivalent to the condition that the velocity of the orbit is much
less the escape velocity from the earth surface.

\section{Acknowledgment}
We thank Benjamin Bromley for pointing out 
the low-velocity interpretation,
given in Eq.~(\ref{lowveloc}), for the
parabola condition.
This work has been partially supported by  the National 
Science Foundation under grant 
PHY0244605.



\begin{thebibliography}{99}
\bibitem{notunique}
Our choice is not the only reasonable way of carrying out this
mapping.  We could alternatively have chosen $x=r\phi$, for
example. By comparison, our choice suffers from having unequal
physical distances, for equal increments of $x$, along $x$ coordinate
lines. Our choice, however, gives much simpler dynamical equations at
the end of this article. The choice made has no distinguishable effect
on the appearance of the results given in the figures.
\bibitem{loworbit} This sort of high velocity motion is not typical 
of practical ballistic trajectories, in which projectiles are launched
at an angle near 45$^o$. For such trajectories $\alpha $ is of order
unity, and the parabolic approximation is accurate. For a given range
and a given $\epsilon$, a larger value of $\alpha$ means a larger
value of $V_{horiz}$, or equivalently of the required muzzle
velocity. Though the conclusion from Eq.~(\ref{range}) is meant to be
a point of principle, not of practical importance, it is interesting
that large guns exist that are able to achieve muzzle velocities of
several km/sec necessary to have peak height $h$ and range, of the
same order as the values $h\sim 1$\,km and range $\sim100$\,km in our
example.
\bibitem{curvature_ellipse} The radius of curvature of the ellipse at its
apogee is $${\cal R}=h\alpha \frac{(1+\epsilon)^2}{1-\alpha\epsilon
(1+\epsilon)}.$$ For $\alpha\epsilon\ll 1$ (which is equivalent to
${\cal R}/R\ll 1$), and for $\epsilon\ll 1$, this coincides with
Eq.~(\ref{curvature_parabola}) at the apogee.
\bibitem{latus_rectum} 
The condition that the parabolic trajectory is a good approximation for 
the true orbit, i.e., that ${\cal R}/R\ll 1$, is equivalent to $L/R\ll 1$,
$L$ being the semilatus rectum.
\bibitem{eqsofmotion} 
These equations can be derived by substituting $r=y+R$ and
$\phi=x/(R+h)$ in the standard equations for central force motion. A
rather simpler equivalent set of equations follows, of course, by
writing the two constants of motion, energy and angular momentum, in
terms of $x,y$ coordinates.  The form in Eq.~(\ref{trueaccel}), with
second derivatives with respect to time, is given to emphasize the
velocity dependence of the acceleration.
\bibitem{notG} One could claim that we have not shown that ``gravity'' 
fails to be vertical. The velocity-dependent terms in
Eqs.~(\ref{trueaccel}), after all, do not contain the gravitational
constant $G$. Those terms are actually kinematic terms due to the 
nature of the coordinates. To avoid this criticism, the fallacious argument 
at the end of Sec.~\ref{sec:intro} was made for
``a particle being acted upon only by gravitational
forces.''
\bibitem{neglect_vel} Notice that the velocity dependent term on the right
hand side of Eq.~(\ref{trueaccel}) is made negligible with respect to  the
term not depending on the velocity if 
$(\,dx/\,dt)/V_{horiz}\ll (\alpha\epsilon )^{-1/2}$, but one cannot
neglect the velocity dependent right hand side of
Eq.~(\ref{trueaccel_x}). Notice, however, that typically indeed 
$(\,dx/\,dt)/V_{horiz}\ll (\alpha\epsilon )^{-1/2}$ and 
the right hand side of Eq.~(\ref{trueaccel_x}) is rather small, such that
Galileo's picture of separate horizontal and vertical motions is a fair
description of the motion. However, for the values of the parameters we
considered above, $\alpha\epsilon$ is about unity, such that it is easy 
to see that the equations of motion (\ref{trueaccel_x}) and
(\ref{trueaccel}) are indeed coupled, and ``gravity is not vertical.''
\end{thebibliography}
\end{document}